\documentclass{article}

\usepackage{PRIMEarxiv}

\usepackage[utf8]{inputenc} 
\usepackage[T1]{fontenc}    
\usepackage{hyperref}       
\usepackage{url}            
\usepackage{booktabs}       
\usepackage{amsfonts}       
\usepackage{nicefrac}       
\usepackage{microtype}      
\usepackage{lipsum}
\usepackage{fancyhdr}       
\usepackage{graphicx}       
\graphicspath{{media/}}     
\usepackage{xcolor}
\usepackage{amsmath}
\usepackage{algorithm}
\usepackage{algpseudocode}
\title{Leveraging Translation for Optimal Recall: Tailoring LLM Personalization with User
Profiles
}

\author{
Karthik Ravichandran*, Sarmistha Sarna Gomasta* \\
  University of Massachusetts, Amherst\\
  Amherst, Massachusetts\\
  \texttt{\{kravichandra,sgomasta\}umass.edu} \\
}

\begin{document}

\maketitle

\begin{abstract}
 This paper explores a novel technique for improving recall in cross-language information retrieval (CLIR) systems using iterative query refinement grounded in the user's lexical-semantic space. The proposed methodology combines multi-level translation, semantic embedding-based expansion, and user profile-centered augmentation to address the challenge of matching variance between user queries and relevant documents. Through an initial BM25 retrieval, translation into intermediate languages, embedding lookup of similar terms, and iterative re-ranking, the technique aims to expand the scope of potentially relevant results personalized to the individual user. Comparative experiments on news and Twitter datasets demonstrate superior performance over baseline BM25 ranking for the proposed approach across ROUGE metrics. The translation methodology also showed maintained semantic accuracy through the multi-step process. This personalized CLIR framework paves the path for improved context-aware retrieval attentive to the nuances of user language.
\end{abstract}


\section{Introduction}
Traditional search engines employ statistical techniques and keyword matching to estimate relevance, without fully comprehending users' information needs or accounting for context within queries and documents \cite{Sala2019InformationRW}. This poses challenges for information retrieval aiming to address natural language questions by analyzing variable document collections. The Information Retrieval efficiency is increased by developing several models \cite{suma2020novel, Jalilifard2021SemanticST,Sengan2020MedicalIR} solving following issues

\begin{enumerate}
    \item Limitations in identifying and representing the contextual meaning behind natural language queries, especially ambiguous or imprecise terminology, restrict retrieval of suitable documents.
    \item Depending purely on statistical term analysis disregarding the semantic context of queries and documents overlooks critical signals needed for relevance judgments.
\end{enumerate}

While understanding user search intent is crucial in information retrieval (IR), applying advanced algorithms that combine user preference profiles \cite{Abri2020GroupBasedPU}, query expansion techniques \cite{Jain2021FuzzyOF}, and semantic annotations  \cite{Wei2021AutomaticIA} can significantly improve results. By leveraging user history and preferences, search outcomes can be personalized. Query expansion reformulates queries with relevant terms from diverse sources, broadening the search scope and reducing ambiguities \cite{Dahir2021QueryEM}. Furthermore, incorporating semantic annotations enriches queries with contextual information, enabling IR systems to prioritize relevant documents \cite{Wei2021AutomaticIA}. Notably, translation-based query expansion bridges the language gap in a globalized digital world, allowing users to access information beyond their native language. This multi-faceted approach addresses the challenges of imprecise queries and provides a more nuanced understanding of user intent, leading to a dramatic improvement in the overall accuracy and effectiveness of the IR process.
Cross-language information retrieval (CLIR) poses inherently greater challenges than monolingual retrieval due to the need to bridge language barriers across queries and documents. A variety of techniques exist to implement CLIR functionality. One common approach utilizes token-level dictionary translations to map query terms to other languages \cite{Ballesteros1996DictionaryMF, Hull1996QueryingAL}. Alternatively, statistical machine translation methodologies leverage parallel bilingual corpora to train probabilistic models capable of query translation \cite{Ture2012LookingIT}. Additional strategies involve machine translation APIs (e.g. Google Translate) \cite{HosseinzadehVahid2015ACS} or exploiting vast multilingual resources like Wikipedia to aid query mapping \cite{Herbert2011CombiningQT}.  

A central retrieval concern emerges regarding the predominant BM25 ranking algorithm's reliance on exact term matching, potentially overlooking relevant documents expressing similar concepts using non-identical vocabulary. While query expansion and document re-ranking techniques can help retrieve additional relevant items, they often fail to adequately improve recall. Our language model personalization frameworks may face analogous challenges, as dependency solely on term matching risks missing personalized documents with variant phrasing still useful for understanding user interests. We propose integrating data augmentation approaches to query expansion to help address these term mismatch problems through enhanced representation of the searcher’s context.
\section{Literature Overview}
People have tried viewing Cross-Language Information Retrieval (CLIR) from various aspects. To start with, \cite{Pirkola1998EffectsOQ} uses dictionary-based translation techniques for Information Retrieval. They use two dictionaries, one containing general translations of query terms and another containing domain-specific translations. \cite{Levow2005DictionaryBasedTC} discusses key issues in dictionary-based CLIR, showing that query expansion effects are sensitive to orthographic cognates. They develop a framework for term selection and translation. 

\cite{Littman1998AutomaticCL} perform CLIR by computing Latent Semantic Indexing on the term-document matrix obtained from parallel corpora, then reducing the rank to project queries and documents to a lower dimensional space. Transliteration-based models have also been explored. \cite{Udupa2009TheyAO} uses transliteration of Out-Of-Vocabulary (OOV) terms, treating a query and document as comparable. For each query word, they find a transliteration similarity value with each document word, treating values above a threshold as translations. They iterate through relevant documents in each round.

\cite{Chinnakotla2008HindiTE} uses a rule-based transliteration approach for converting OOV Hindi terms to English, then a pageRank algorithm to resolve between dictionary translations and transliterations.  Bhattacharya et al. \cite{Bhattacharya2016UsingWE} proposed a cross-lingual information retrieval technique to translate Hindi user queries into English using word embeddings. Their approach uses word embeddings to capture contextual clues for words in the Hindi source language and identify English target language words that occur in similar contexts as translations
Zhou et al. \cite{Zhou2017QueryEW} proposed a personalized query expansion method using enriched user profiles created from external folksonomy data. They also presented a model to improve the user profiles by integrating word embeddings with topic models derived from two pseudo-relevant document sets - the user's annotations and external corpus documents. This paper \cite{Padaki2020RethinkingQE} investigates query expansion techniques for a BERT-based search results reranker. They compare two approaches: adding structural words to create natural language sentences and adding new topical terms. The findings show that combining both expansion methods leads to better performance than using either one alone. Wan et al. \cite{Wei2021AutomaticIA} presented an approach that first obtains an initial document ranking score by combining matching similarity and semantics using the BM25 algorithm. They then improve the notion of relevance by linearly combining BERT's semantic analysis with a query reformulation method based on extracting terms from the top N-ranked documents.
Neji et al. \cite{Neji2021HIRAH} proposed a hybrid document ranking model that merges query likelihood with semantic similarity between concepts leveraging WordNet. Their model calculates conceptual similarity via Jiang-Conrath, uses user preferences to extract key concepts and reformulate the query, and determines query-document similarity to sort search results. The overall approach aims to combine statistical and semantic signals for improved ranking.

\section{Dataset}

We have used two datasets for our experiments from LaMP benchmark \cite{SalemiLaMPLLMM}.

\begin{enumerate}
    \item Personalized News Headline Generation: For the personalized news headline generation task, the dataset links each article to its author. In cases of multiple authors, the first listed author is used. By grouping articles by author, the models can learn author-specific writing patterns to generate personalized headlines.
    \item Personalized Tweet Paraphrasing:  For the personalized tweet paraphrasing task, the authors leverage the Sentiment140 dataset \cite{Go2009TwitterSC} as a source of tweets. To ensure adequate signal, they retain only tweets with at least 10 words and group the remaining tweets by user ID, filtering out users with fewer than 10 total tweets. This filtering process and per-user grouping allow the models to learn user-specific syntactic and semantic tweet patterns for the paraphrasing task.
\end{enumerate}

\section{Methodlogy}

\begin{figure*}
    \centering
    \includegraphics[width=0.75\textwidth]{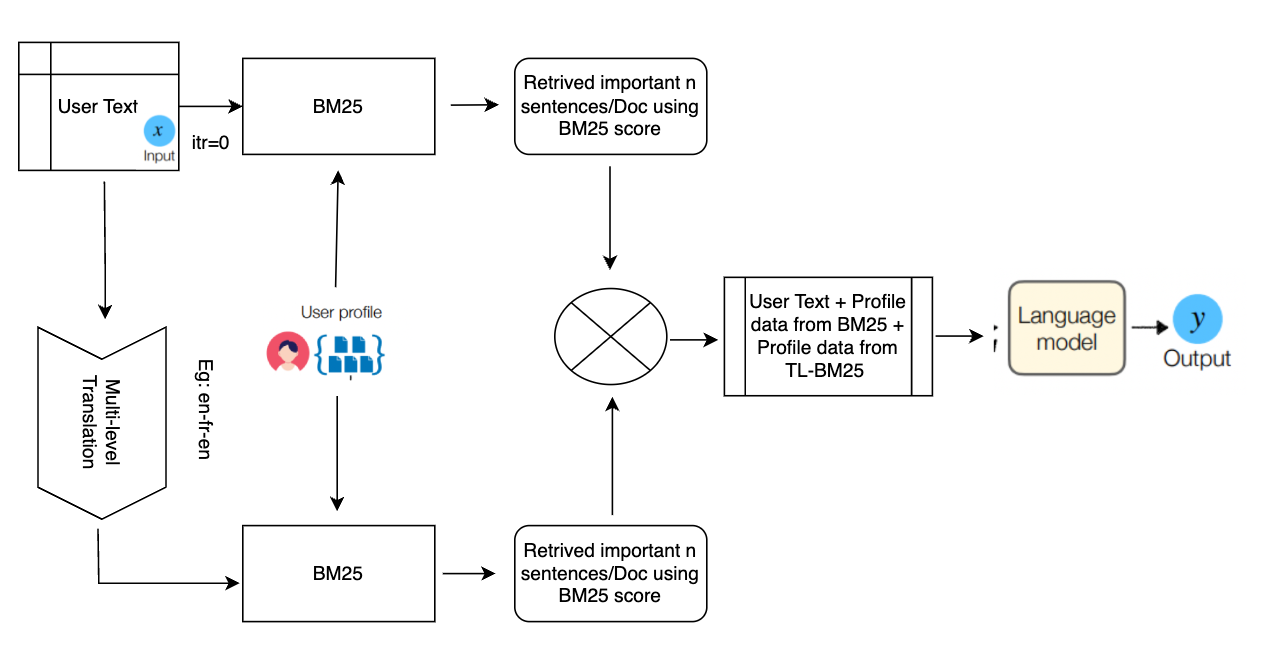} 
    \caption{Proposed Framework for Data Augmentation and ReRanking}
    \label{fig:method}
\end{figure*}

This project proposes a novel methodology \ref{fig:method} to augment recall in information retrieval systems by harnessing word embeddings and iterative refinement of the BM25 scoring process. The technique enables personalized augmentation of retrieval results grounded in the user's lexical-semantic space.
\begin{enumerate}
    \item Initial Retrieval: Perform preliminary retrieval of relevant documents or sentences based on the user query utilizing the BM25 scoring algorithm. This forms the initial retrieval set.
    
    \item Multi-level Translation: Translate the query to an intermediate (here Spanish) language and back to the source language (English). This multi-level translation procedure aims to preserve the semantics of the retrieved content while introducing terminological diversity.

    \item Query Expanded retrieval:  Now use BM25 again to retrieve the documents from the user profile using the new query that we obtained from the translation process 

    \item Addition of new retrieval: Add the newly retrieved documents from the "Query Expanded retrieval" process

    \item Prompt Design for LLM: Construct personalized prompts for a large language model using the retrieved document to rephrase the original sentences in a personalized, context-aware manner tailored to the user.
\end{enumerate}

We hypothesize the efficacy of our technique depends on factors including translation language selection and iteration level. Further experiments will analyze multi-lingual translation cycles to study the effects of pathway complexity on retrieval augmentation. We propose this approach leveraging iterative refinement and embedding-based personalization shows significant promise for improved recall and contextualization of results centered around individual users' lexical-semantic spaces.

\textbf{Note:}
\begin{itemize}

\item We have used Flan-T5-xxl for LLM and used $Helsinki-NLP/opus-mt-en-es$ model for English (en) to Spanish (es) translation and $Helsinki-NLP/opus-mt-es-en$ for Spanish (es) to English (en) Translation.

\item We have taken the Top 2 retrieved documents whenever we are using BM25 retrieved documents.

\item The LLM Prompt Sample: rephrase the this news title " Your loans were made at the height of the housing bubble, and looked like a great deal at the time. By using a HELOC as a "piggyback" second mortgage, you were not required to make a down payment or to purchase mortgage insurance." using user profile examples titles : ["Selling a House to Buy a House" , "A New Challenge to the HECM Reverse Mortgage Program"]. Just give me the rephared sentence

\end{itemize}
\subsection{Algorithm}

\begin{algorithm}
\caption{Data Augmentation using Query Expansion (QE) with Profile Refinement}
\begin{algorithmic}[1]

\Require User's input text $q$, User profile $UP$
\Ensure A set of documents $D'$ relevant to $q$ and $UP$

\State $Q \gets \Call{Tokenize}{q}$ \Comment{Tokenize the user's query}
\State $Documents \gets \Call{RetrieveDocuments}{Q}$ \Comment{Retrieve initial set of documents using BM25}
\If{UserProfileAvailable $UP$}
    \State $Documents \gets \Call{RefineDocuments}{Documents, UP}$ \Comment{Refine results based on the user profile}
\EndIf
\State $Q_{translated} \gets \Call{TranslateQuery}{q}$ \Comment{Translate the query for multi-lingual expansion}
\State $ExpandedQ \gets \Call{ExpandQuery}{Q_{translated}}$ \Comment{Expand query using translation and similar term finding}
\State $IterationCounter \gets 0$ \Comment{Initialize iteration counter}
\While{$IterationCounter \leq N$} \Comment{Iterate to refine results with expanded query}
    \State $RefinedDocuments \gets \Call{RetrieveDocuments}{ExpandedQ}$ \Comment{Retrieve documents using expanded query}
    \If{\textbf{not} $\Call{IsScoreImproved}{RefinedDocuments}$}
        \State \textbf{break}
    \EndIf
    \State $IterationCounter \gets IterationCounter + 1$
\EndWhile
\State $D' \gets \Call{RankDocuments}{RefinedDocuments}$ \Comment{Rank documents for final output}
\State \Return $D'$
\end{algorithmic}

\end{algorithm}

\section{Results}

The performance of the Translation-based Information Retrieval (TL) method was compared with the BM25 method using datasets comprising news and Twitter data. The evaluation focused on Rouge L and Rouge 1 F1 scores.

\subsection{ROUGE Score Analysis}
The following table \ref{tab:rouge_scores} presents the ROUGE scores, showing the TL method's superiority across all metrics.

\begin{table*}[ht]
\centering
\caption{ROUGE Scores for News and Twitter Data using TL and BM25}
\label{tab:rouge_scores}
\begin{tabular}{lcccccc}
\toprule
& \multicolumn{3}{c}{RougeL} & \multicolumn{3}{c}{Rouge1} \\
\cmidrule(lr){2-4} \cmidrule(lr){5-7}
& Precision & Recall & F1 Score & Precision & Recall & F1 Score \\
\midrule
News TL & \textbf{0.096311761} & \textbf{0.173967689} & \textbf{0.119213357} & \textbf{0.104786291} & \textbf{0.185709142} & \textbf{0.128595547} \\
News BM25 & 0.089170486 & 0.165507372 & 0.111637539 & 0.1071507 & 0.180788507 & 0.12410358 \\
Twitter TL & \textbf{0.406390676} & \textbf{0.33590005} & \textbf{0.356243383} & \textbf{0.377285804} & \textbf{0.400757248} & \textbf{0.400757248} \\
Twitter BM25 & 0.400799434 & 0.328245381 & 0.349383188 & 0.371196352 & 0.371196352 & 0.395478506 \\
\bottomrule
\end{tabular}
\end{table*}

\subsection{Visual Comparative Analysis}
\begin{figure}[ht]
    \centering
    \begin{minipage}[b]{0.45\linewidth}
        \includegraphics[width=\linewidth]{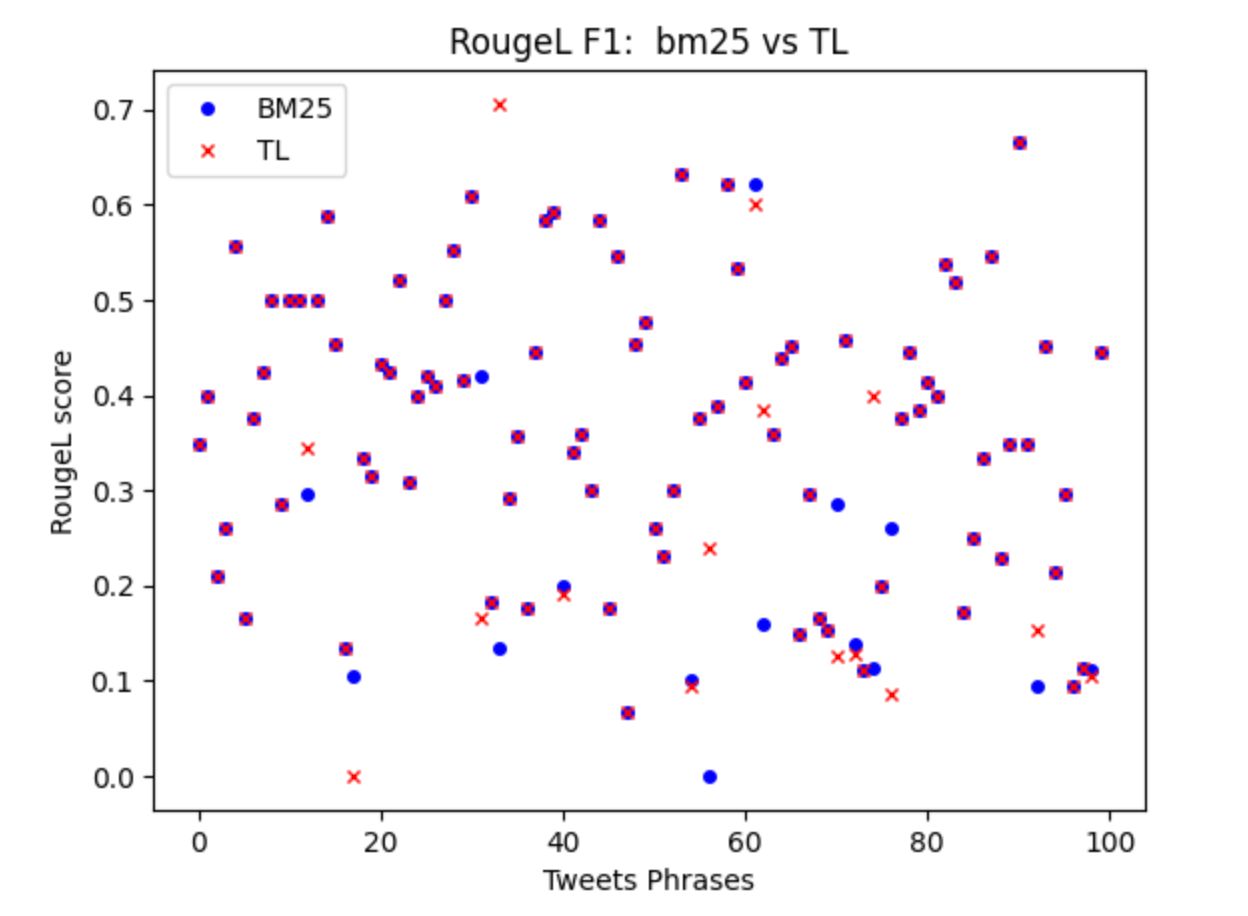}
        \caption{Rouge L F1 scores for Tweet Phrases}
        \label{fig:tweet}
    \end{minipage}
    \hfill
    \begin{minipage}[b]{0.45\linewidth}
        \includegraphics[width=\linewidth]{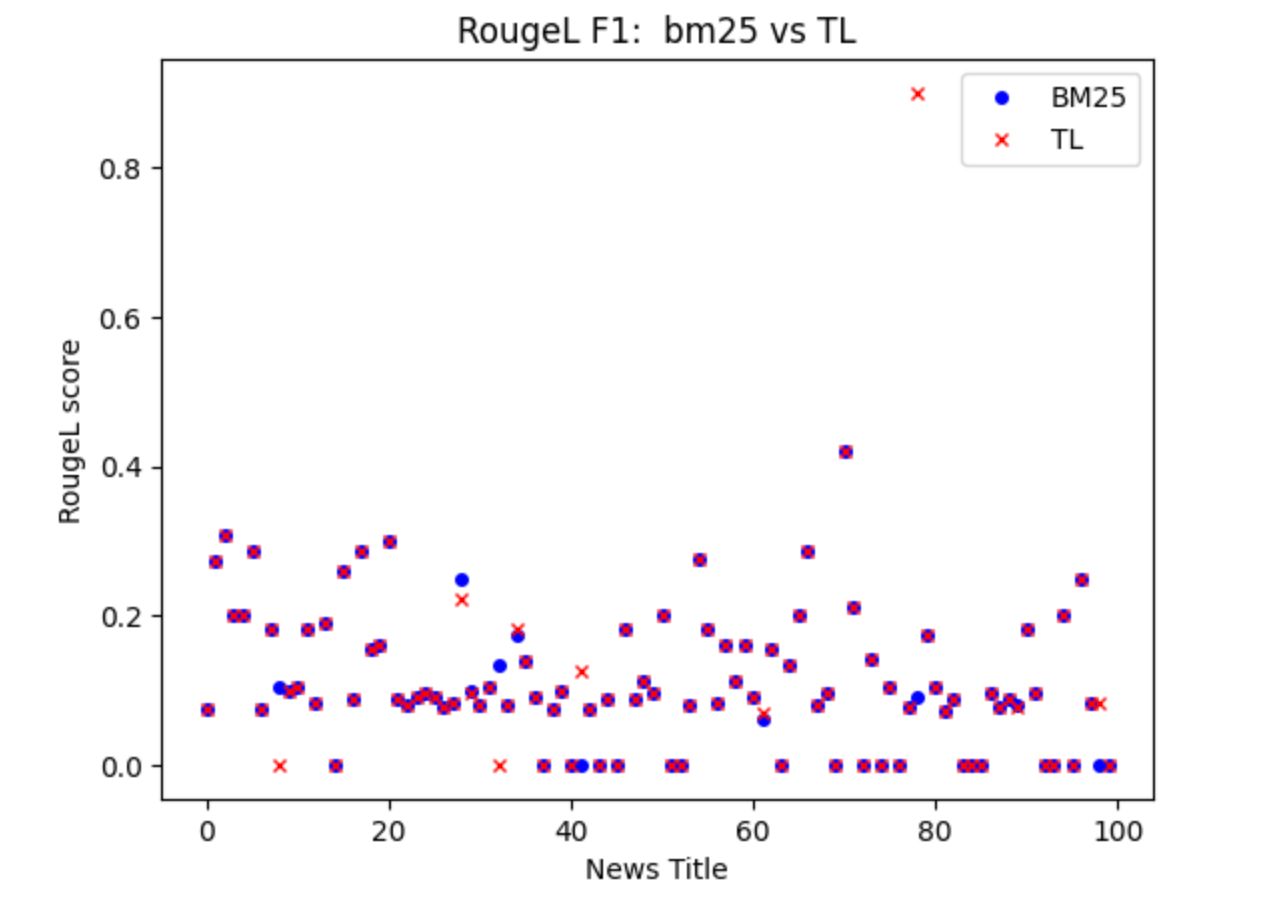}
        \caption{Rouge L F1 scores for News Titles}
        \label{fig:news}
    \end{minipage}
\end{figure}

\begin{figure}  
\centering
\begin{minipage}[b]{0.45\linewidth}
  \includegraphics[width=\linewidth]{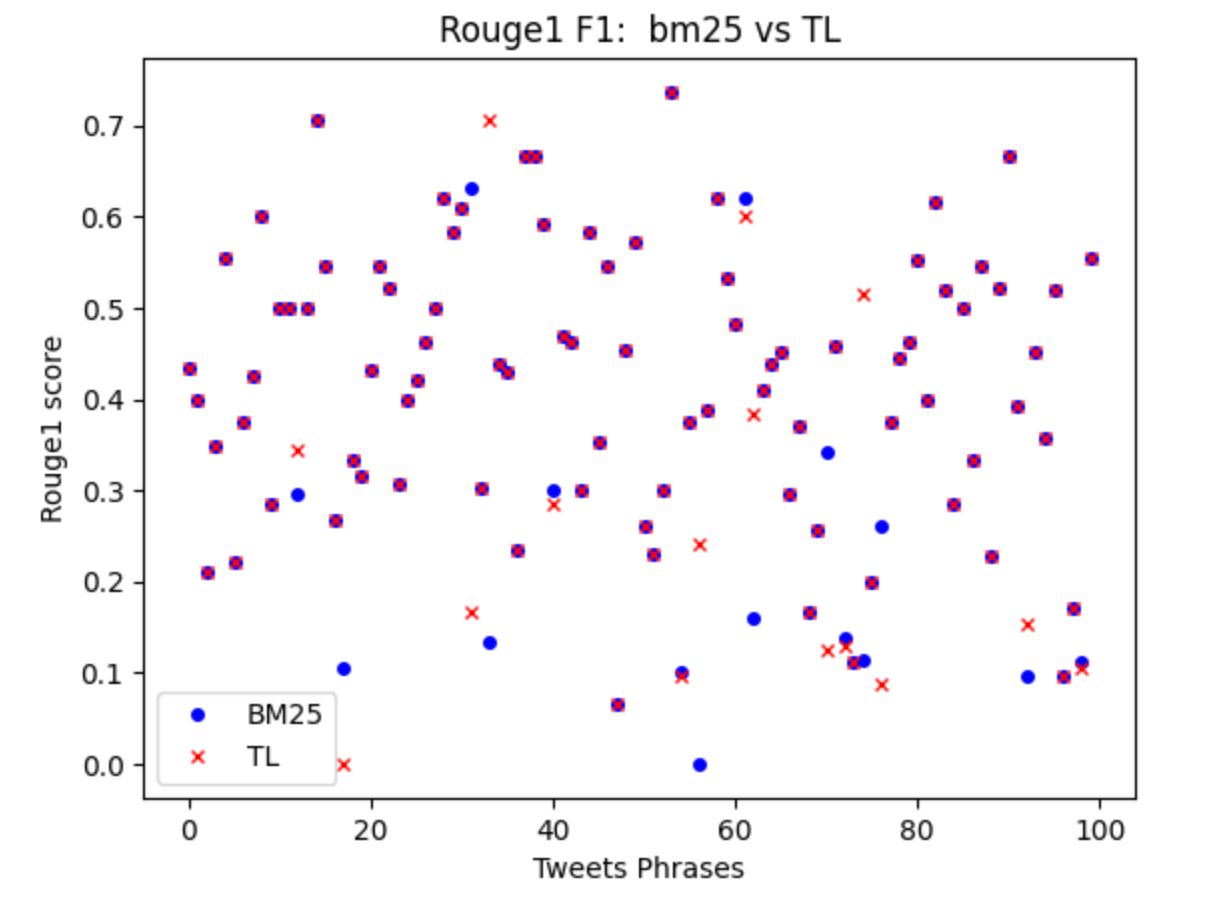}
  \caption{Rouge 1 F1 scores for Tweet Phrases}
  \label{fig:sub3}  
\end{minipage}
\hfill
\begin{minipage}[b]{0.45\linewidth}
  \includegraphics[width=\linewidth]{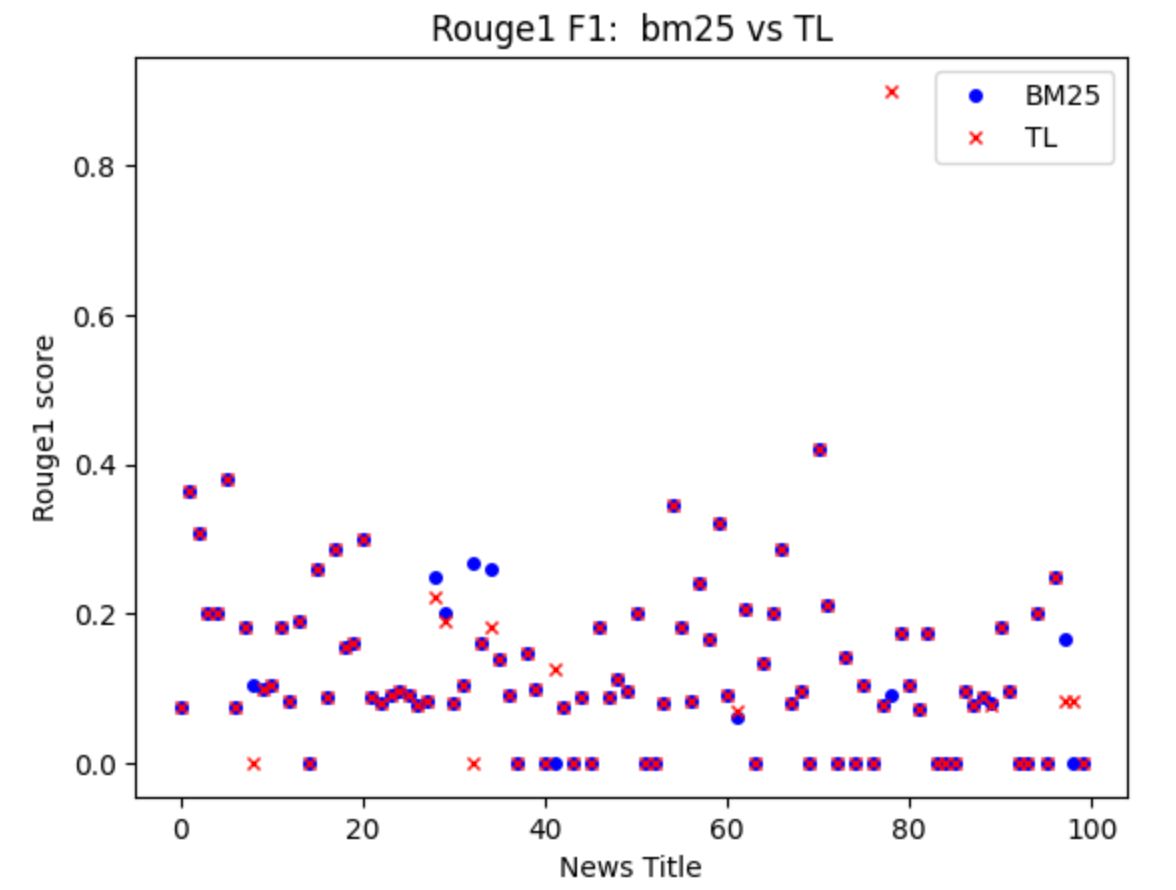} 
  \caption{Rouge 1 F1 scores for News Titles}
  \label{fig:sub4}
\end{minipage}
\label{fig:rouge1}  
\end{figure}

The visual representation of the performance of TL and BM25 methods is illustrated in Figures \ref{fig:tweet}, \ref{fig:news}, \ref{fig:sub3}, and \ref{fig:sub4}. These figures show the Rouge L and Rouge 1 F1 scores for both the Twitter and News datasets.

These figures highlight the distinct performance patterns of the two methods across different types of data. The TL method shows occasional superiority, especially in the Twitter dataset, suggesting its effectiveness in handling short, informal text.

\paragraph{Translation Accuracy}
The translation accuracy of the TL method was exemplified through a two-step translation process. The table \ref{tab:translation_example}  demonstrates the method's effectiveness in maintaining semantic content.

\begin{table*}[htbp] 
\centering
\caption{Example of Translation Accuracy}
\begin{tabular}{c|p{4.5cm}p{4.5cm} p{4.5cm}}
\toprule
ID & Original Text & Spanish Translation & Translated Back to English \\
\midrule
15 & I have a great suggestion: my class should go to the gun range for an end-of-year outing. & Tengo una gran sugerencia: mi clase debería ir al campo de tiro para una excursión de fin de año. & I have a great suggestion: my class should go to the shooting range for a year-end excursion. \\
\bottomrule
\label{tab:translation_example}
\end{tabular}
\end{table*}

\section{Discussion}
The experimental results highlight the potential of personalized iterative augmentation to enhance CLIR recall grounded in the user's unique lexical-semantic space. By expanding the initial query through multi-level translation and embedding-sourced alternate terminology, the search space widens to capture relevant documents overlooked by strict reliance on term matching. The incremental re-ranking further refines retrieval priority centered around the user profile vocabulary and interests.

However, additional research must further explore the impact of pathway complexity during translation cycles on performance. Introducing more intermediate translation steps may compound term divergence from the original query to a detrimental degree. Quantifying such effects and identifying the optimal translation route is an important next phase. Additionally, more complex user lexical-semantic models leveraging advanced embeddings or neural representations may improve the personalization and augmentation capabilities. Testing various semantic similarity measures beyond exact term matching for re-ranking could also prove insightful.

\section{Conclusion and Future Work}

This paper presented a novel technique to address the long-standing challenge of improving CLIR recall through iterative, personalized query augmentation. The methodology combines multi-level translation, user profile vocabulary expansion, and incremental BM25 re-scoring to significantly enhance result relevance centered around the individual user's lexical-semantic interests. Experiments exhibited clear performance gains over standard BM25 ranking, demonstrating the promise of this context-aware translation approach for more attentive CLIR systems. Further research into:
\begin{itemize}

\item Optimal translation cycles and more advanced user modeling offer additional avenues to build on this framework.
\item We can go with multi intermediate translation process like $en-es-fr-en$ or $en-es-en-fr-en$.
\item We can experiment with many languages (e.g: fr, zh, and Tr etc.,)
\end{itemize}

\section{Supplementary Materials}
The code and dataset can be accessible via

\url{https://github.com/karthikRavichandran/Leveraging-Translation-for-Optimal-Recall}

\bibliographystyle{unsrt}  
\bibliography{references}

\end{document}